\documentclass{aa}

\usepackage{graphicx}
\usepackage{txfonts}
\usepackage{float}
\usepackage{amssymb} 

\begin{document} 

\title{Photometry of \emph{K2} Campaign 9 bulge data}

\author{
R.~Poleski\inst{1,2} \and 
M.~Penny\inst{1} \and 
B.~S.~Gaudi\inst{1} \and 
A.~Udalski\inst{2} \and 
C.~Ranc\inst{3} \and 
G.~Barentsen\inst{4,5} \and 
A.~Gould\inst{6,7,1}
} 

\institute{
Department of Astronomy, Ohio State University, 140 W. 18th Ave., Columbus, OH 43210, USA 
\\ \email{poleski.1@osu.edu}
\and Warsaw University Observatory, Al. Ujazdowskie 4, 00-478 Warszawa, Poland 
\and Astrophysics Science Division, NASA/Goddard Space Flight Center, Greenbelt, MD 20771, USA
\and NASA Ames Research Center, Moffett Blvd, Mountain View, CA 94035, USA
\and Bay Area Environmental Research Institute, 625 2nd St., Ste. 209, Petaluma, CA 94952, USA
\and Korea Astronomy and Space Science Institute, 776 Daedukdae-ro, Yuseong-gu, Daejeon 34055, Korea
\and Max-Planck-Institute for Astronomy, K\"{o}igstuhl 17, 69117 Heidelberg, Germany
}

\date{Received ???; accepted ???}

\abstract{
In its Campaign 9,  \emph{K2} observed dense regions toward the Galactic bulge in order
to constrain the microlensing parallaxes and probe for free-floating planets. 
Photometric reduction of the \emph{K2} bulge data poses
a significant challenge due to a combination of the very high stellar density, large pixels of the \emph{Kepler} camera,
and the pointing drift of the spacecraft.
Here we present a new method to extract \emph{K2} photometry in dense stellar regions. 
We extended the Causal Pixel Model developed for less-crowded fields, first by  
using the pixel response function together with accurate astrometric grids, 
second by combining signals from a few pixels, and third by 
simultaneously fitting for an astrophysical model.
We tested the method on two microlensing events and a long-period eclipsing binary.  
The extracted \emph{K2} photometry is an order 
of magnitude more precise than the photometry from other method. 
}


\maketitle

\section{Introduction}

In May 2013, the original \emph{Kepler} mission \citep{borucki10} suffered from the failure of the second of its four reaction wheels 
and hence lost the ability to maintain stable pointing.  
To compensate for spacecraft pointing drifts, the \emph{Kepler} satellite used 
solar radiation pressure to partially stabilize its pointing.  
The sacrifice of this solution is that \emph{Kepler} can only look at 
locations near the ecliptic plane, and then only for roughly 80 days.  
Further, there is still a residual pointing drift 
with an amplitude of about one pixel
over a period of 6.5 hours.  Thus, the photometric signals are
dominated by sensitivity variations of the detector on the subpixel scale as
bright stars drift across the pixels.  
The repurposed mission was named
\emph{K2} \citep{howell14} and observed Ecliptic fields in a series of 
$\approx80$-day-long campaigns.  One of these was
\emph{K2} Campaign 9 (\emph{K2}C9), which was devoted to a microlensing experiment
\citep{gould13b,henderson16}.  In \emph{K2}C9 almost all pixels available for downlink 
were selected in a nearly continuous superstamp \citep{henderson16}, 
which made \emph{K2} the first wide-field microlensing survey carried out by a satellite.  
The most important capability of \emph{K2}C9 was to directly measure 
masses of microlenses without requiring target selection, which, in principle, enabled mass measurements of free-floating planets for 
the first time \citep[][see method description below]{penny17}.  
The first estimate from microlensing of an occurrence rate of a free-floating planet was very high and was based 
on the distribution of event timescales, meaning 
the occurrence rate was inferred indirectly \citep{sumi11}.  
The events of short duration can be caused not only by free-floating planets 
but also by planets on very wide orbits \citep{hand05d}, 
both of which are difficult to study and scientifically important \citep{poleski14c,mroz18a}. 
Simultaneous observations of short-timescale events from the ground 
and from a satellite directly constrain the lens mass \citep{refsdal66,gould94c} 
and hence verify that the observed short-timescale events 
are due to planetary-mass objects.  
The mass is measured directly if we can measure the Einstein ring radius ($\theta_{\rm E}$) 
and the microlensing parallax ($\pi_{\rm E}$):
\begin{equation}
M = \frac{\theta_{\rm E}}{\kappa \pi_{\rm E}},
\end{equation}
where $\kappa = 4G/(c^2{\rm AU}) = 8.14~{\rm mas~M_{\odot}^{-1}}$. 
The microlensing parallax vector can be measured by comparing ground-based and satellite 
impact parameters ($u_0$) and epochs of closest approach ($t_0$):
\begin{equation}
\mbox{\boldmath$\pi$}_{\rm E} \approx \frac{\rm AU}{D_\perp}\left(\frac{t_{0,{\rm sat}}-t_{0,\oplus}}{t_{\rm E}},\pm u_{0,{\rm sat}}\mp u_{0,\oplus}\right),
\end{equation}
where $t_{\rm E}$ is Einstein timescale and $D_{\perp}$ is the Earth--satellite separation 
projected on the sky.
Measurement of lens masses in the shortest timescale events 
cannot be obtained by employing \emph{Spitzer}, the other satellite used 
for microlensing parallax measurements,  due to the small field-of-view of its camera 
and scheduling requirements \citep{yee15b} that favor 
the observations of medium-length and longer events.  
During \emph{K2}C9, the superstamp was intensively
observed from the ground \citep{henderson16} and no short-timescale events 
($t_{\rm E}< 2~{\rm d}$) were detected. It is possible that this is partially due to unusually bad weather during the \emph{K2}C9
at Chilean observatories, which would have contributed a significant part of the microlensing data. 
After \emph{K2}C9, \citet{mroz17b} analyzed a few years of 
the high-cadence observations by 
the Optical Gravitational Lensing Experiment (OGLE) 
and demonstrated that the rate of short events 
is much smaller than previously claimed.  A decrease in the expected number of 
short events significantly reduced the interest in photometric reduction 
of \emph{K2}C9 data, which was recognized  early on to be 
a very challenging task. 

The original \emph{Kepler} mission 
produced highly accurate photometry 
thanks to stable pointing and a low density of stars.  
There are a number of aspects that make \emph{K2} photometry of bulge fields difficult:
the spacecraft pointing is not stable; the pixel scale is large; 
the pixel response function (PRF) is undersampled, yet is extended and varies across the field; 
and the bulge fields have an extremely high density of stars.  
A combination of all these factors produced a data set that is considerably more difficult to analyze 
than would be the case for a data set affected by only one of these aspects.

\begin{figure}[!h]
\centering
MOA-2016-BLG-290
\medskip
\includegraphics[width=.96\hsize]{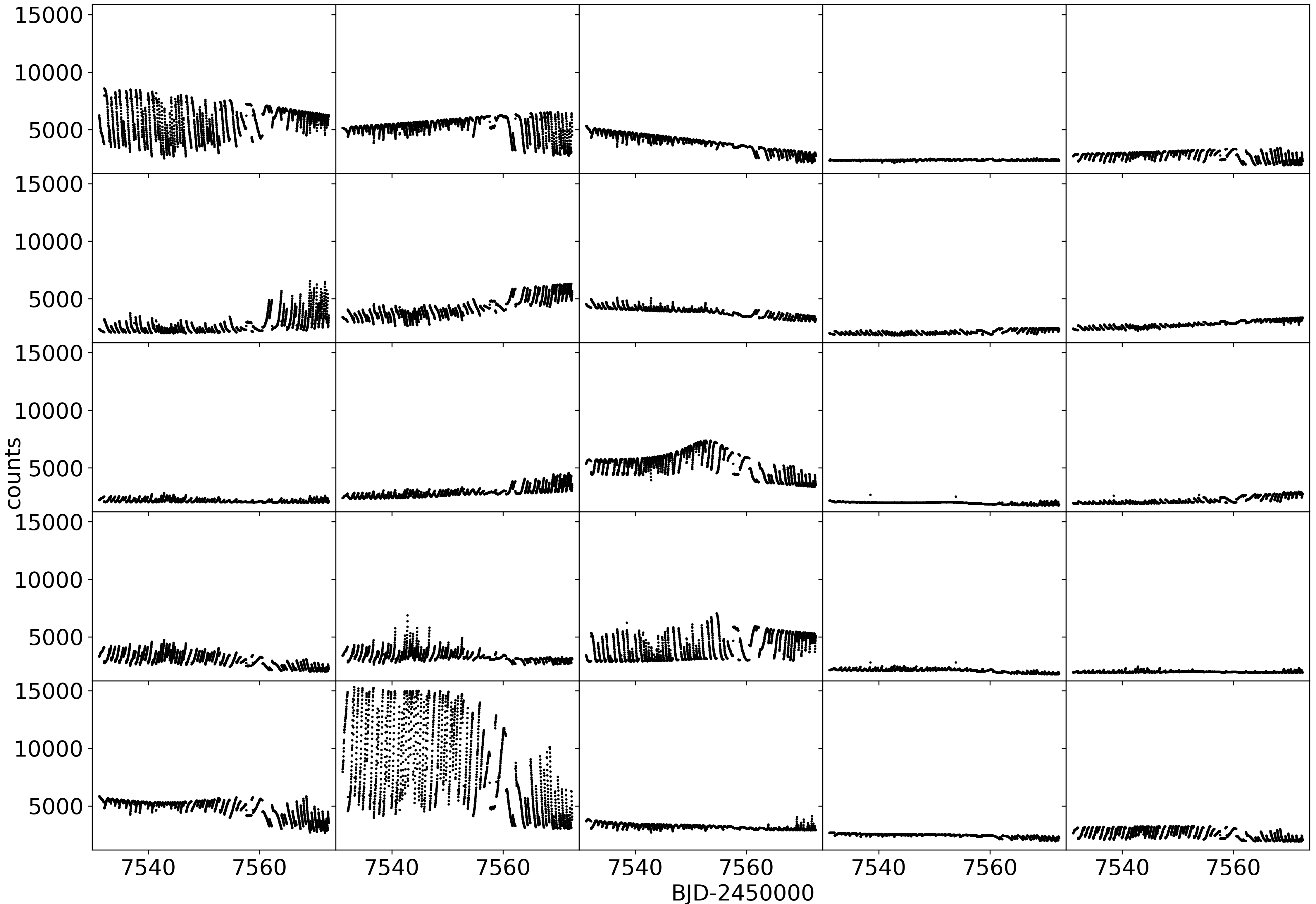} 

\bigskip\bigskip\bigskip\bigskip
OGLE-2016-BLG-0795
\medskip
\includegraphics[width=.96\hsize]{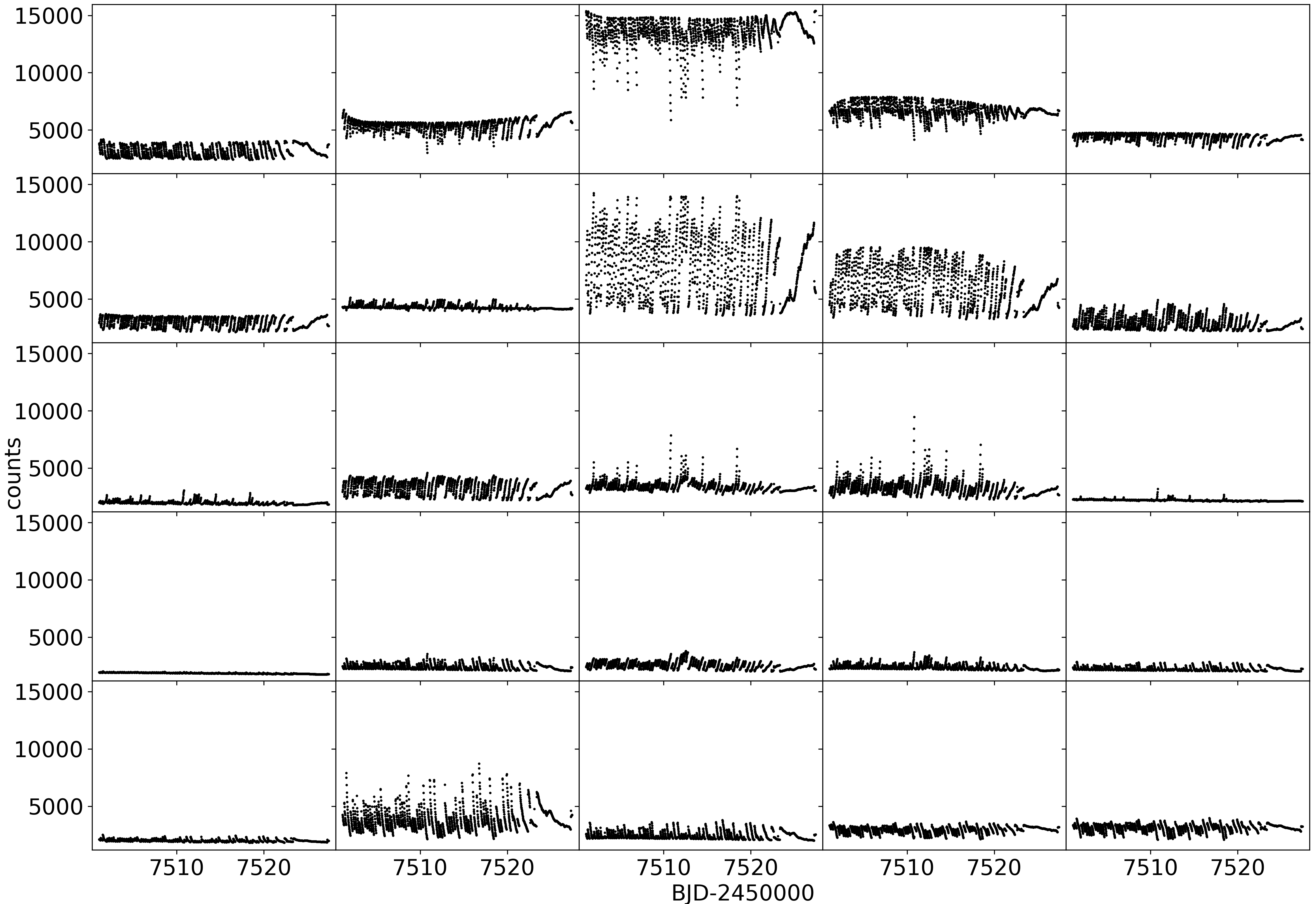}
\caption{Raw light curves in 25 adjacent pixels centered on very bright 
microlensing event MOA-2016-BLG-290 (top, C9b) and significantly fainter event OGLE-2016-BLG-0795 (bottom, C9a). 
The Y-axis values (in $\mathrm{e^-}\,s^{-1}$) are very precisely measured: 
the uncertainties are in the range 1.8-3.6 counts, i.e., invisible on the plot above. 
There are 2022 epochs in each panel of the top plot and 1278 in each panel of the bottom plot.
The instrumental trends have patterns that are shared by different pixels. The instrumental trends are larger (and in many cases much larger) 
than the microlensing signal, as the microlensed source is usually 
not the brightest star in a K2 pixel. 
For OGLE-2016-BLG-0795 we extract photometry using the central pixel 
and three pixels adjacent to it (bottom, left, and right).
\label{fig:tpf}}
\end{figure}

Here we introduce the Modified Causal Pixel Model (MCPM) for extraction of 
\emph{K2} bulge photometry.  MCPM is a significant advance upon the Causal Pixel Model (CPM) 
by \citet{wang16}, 
which was developed for photometry of 
planetary transits in less crowded \emph{K2} campaigns.  
The basic idea behind CPM is to remove the instrumental 
trends in the photometry, which are highly correlated between different pixels, see Figure~\ref{fig:tpf}. 
A linear combination of signals 
observed in pixels far from the target is used to model 
the instrumental trends in the target pixel.  

The CPM method \citep{wang16} was designed for planetary transits and 
takes advantage of the fact that transits last 
only a short period of time relative to the full span of the data, and have relatively low amplitudes.  
Most of the time, the target is at the baseline brightness and, therefore, 
there are many epochs that can be used for finding linear dependencies 
between signals observed in different pixels, or training the model.  
In contrast to planetary transits, most microlensing events show significant flux variations 
over long periods of time. Typical Einstein timescales are 
between 10 and 40 days \citep{wyrzykowski15}, 
and significant flux variations can be seen over a few 
$t_{\rm E}$.  In most cases, the event lasts longer than the length of a single 
\emph{K2} subcampaign of around 40 days.  
Hence, only a small fraction of the events have data 
taken over both the baseline and the event peak during the same subcampaign. 
Additionally, there are very few epochs that can be used for training the model.  
This lacuna forces us to  
extract photometry and fit the astrophysical model simultaneously.

The first method of extracting \emph{K2}C9 photometry was presented by 
\citet{zhu17a}.  They further developed the method 
by \citet{huang15}, which was aimed at less crowded \emph{K2} fields.   
In this method, first the difference images are calculated, 
then the aperture photometry is extracted from the difference images, 
and finally this photometry is decorrelated against pointing parameters.  
The decorrelation 
is done simultaneously with 
microlensing model fitting.  
The \citet{zhu17a} method was later used by 
\citet{zhu17c}, \citet{ryu18a}, and \citet{zang18}.  
\citet{libralato16a} have also developed a crowded-field \emph{K2} photometry technique, 
though it has not been applied to \emph{K2}C9 as of yet. 

The challenging nature of extracting \emph{K2} crowded-field photometry, 
and the lack of publicly available tools to do so, have almost certainly 
held back microlensing studies based on the K2 data.  
This work aims to address some of the challenges, and make the  tools for photometry extraction publicly available.

In the next section we present the \emph{K2} bulge data.  
Our method is described in Section~\ref{sec:met}.  
In Section~\ref{sec:ex} we apply our method to a few examples.  
In Section~\ref{sec:flux} we discuss \emph{K2} flux calibration.
We conclude in Section~\ref{sec:end}. 

\section{\emph{K2} bulge data}

The \emph{K2}C9 was divided into two subcampaigns (C9a and C9b),  
with a data downlink during the break in between, in order to 
increase the sky-area surveyed.  This resulted in a superstamp covering 
$3.7~{\rm deg^2}$ \citep{henderson16},  
which was then selected to maximize the observed event rate \citep{poleski16a}.  
The camera field of view was slightly shifted between the subcampaigns.  
The cadence of 
\emph{K2} data was $30~{\rm minutes}$.   
Hence, in subcampaigns C9a and C9b there were 1290 and 2022 epochs collected, 
respectively.  
About $10\%$ of epochs in each subcampaign are affected by spacecraft 
thrusters firing and we exclude these epochs from analysis. 
The pixel scale is $3\farcs98$. 
The \emph{K2} camera is divided into channels of $1100\times1024$ pixels.
The superstamp is at the edge of the camera and 
falls in channels numbered 30, 31, 32, 49, and 52.
The entire channel 31 was within the superstamp, while only sections 
of the other channels were included \citep{henderson16}.
We obtained the \emph{K2} data from the Mikulski Archive for Space Telescopes. 

In addition to the superstamp observations, 
selected events detected by the ground-based microlensing surveys 
\citep[mainly the OGLE Early Warning System;][]{udalski03}
early in the season were scheduled for observations 
\citep[these are called ``late targets'';][]{henderson16}.  
Additional \emph{K2} observations of the bulge were performed in Campaign 11 (C11), 
but in this part of the bulge the event rate is lower.  
Thus, no superstamp was selected in \emph{K2}C11 and only late targets were observed.

\section{Method description}
\label{sec:met}

In order to extract the photometry from \emph{K2} data, the MCPM first assumes 
a model light curve, then uses this model to detrend the signal 
in target pixels, 
and finally combines the detrended signals to extract the photometry. 
The extracted photometry is compared to the assumed model in order 
to calculate $\chi^2$ for a given model. Hence, if the assumed model is 
different than the real signal present in the data, then the resulting $\chi^2$ 
is large. 

In the MCPM 
the  flux $f_{m,i}$ integrated  in pixel $m$ at epoch $i$ is decomposed into 
the astrophysical difference flux and the instrumental trends.  
The astrophysical difference flux in 
any given pixel is the total astrophysical difference flux $\tilde{F}_i$ from the target object 
multiplied by the appropriate value of the PRF.  
The $PRF(x_p-x_*, y_p-y_*)$ is the total flux measured in the pixel centered at 
$(x_p, y_p)$ due to a star with centroid at $(x_*, y_*)$ \citep{anderson00}.  
We describe the estimation of the PRF values in 
detail in Section~\ref{sec:prf} and, for simplicity, index the PRF with pixel ($m$) and epoch ($i$), 
that is, $PRF_{m,i}$.  The second contribution to the signal in a given pixel 
comes from the instrumental trends.  To model the instrumental trends 
the MCPM follows the CPM approach and represents these trends 
as a linear combination of the fluxes 
observed at the same epoch $i$ but in different pixels $m'$, meaning, 
$\sum_{m'}a_{m,m'}f_{m',i}$, where $a_{m,m'}$ are the coefficients that are 
found by fitting as described below and are independent of time.  
The number of pixels used for training ($M'$) is a few hundred.  
Finally, we derived the following equation: 
\begin{equation} \label{eq:main}
\tilde{f}_{m,i} = \tilde{F}_i PRF_{m,i} + \sum_{m'} a_{m,m'}f_{m',i},
\end{equation}
where $\tilde{f}_{m,i}$ is the MCPM estimate of $f_{m,i}$ and can be thought of  
as the model flux for a given pixel.

There are many sets of values of $a_{m,m'}$ that would produce similar 
results in Eq.~\ref{eq:main}. 
We designed the MCPM so that it finds $a_{m,m'}$ values that  describe the data 
well while simultaneously avoiding the danger of overfitting.  
The MCPM follows the approach taken by \citet{wang16} and regularizes 
the system of equations using L2 regularization, meaning
the MCPM adds a term $\lambda\sum_{m'}a_{m,m'}^2$ to the $\chi^2$ 
in order to favor values of $a_{m,m'}$ that are small:
\begin{equation} \label{eq:l2}
\chi^2_m = \sum_i \frac{\left(f_{m,i}-\tilde{f}_{m,i}\right)^2}{\sigma_{m,i}^2} + \lambda\sum_{m'}a_{m,m'}^2,
\end{equation}
where $\lambda$ is the regularization strength. The signal in a pixel $m$ with 
instrumental trends removed is:
\begin{equation}
 \delta f_{m,i} = f_{m,i}-\sum_{m'}a_{m,m'}f_{m',i}.
\end{equation}
The $\chi^2_m$ minimization is run separately for each pixel. 
The simplest approach for finding the astrophysical difference flux is to take 
the sum of the $\delta f_{m,i}$ over $M$ pixels: $\sum_m \delta f_{m,i}$. 
This approach leads to acceptable results, but a more efficient approach 
is to perform a PRF-like photometry and assume that the $\delta f_{m,i}$ 
are already background-corrected.  
The MCPM finds $F_i$ by minimizing the residuals of a system of equations: 
\begin{equation}
\delta f_{m,i} = F_i PRF_{m,i}
,\end{equation}
which leads to:
\begin{equation}
F_i = \frac{\sum_m PRF_{m,i}\delta f_{m,i}}{\sum_m PRF_{m,i}^2}.
\end{equation}

There are hundreds of nuisance parameters in the MCPM, making the model 
very flexible. In turn, this flexibility can affect the fitting convergence.    
Added to this, microlensing model fits suffer from multiple degeneracies.  
In particular, a continuous degeneracy exists between $t_{\rm E}$,  
the source flux ($F_s$), and $u_0$ \citep{wozniak97,han99d}.   
In practice, to reduce \emph{K2} data we must also simultaneously fit  
ground-based data for the same event in order to constrain the timescale and 
possibly also the source flux \citep{zhu17a,zang18} when fitting the microlensing model to 
the \emph{K2} data. For the brightest and the shortest events 
the ground-based data may not be needed.
There are multiple ground-based datasets of the 
\emph{K2}C9 superstamp area that were collected during the campaign and some of them are public: 
the Korean Microlensing Telescope Network \citep[KMTNet;][]{kim18b}, 
the United Kingdom Infrared Telescope \citep[UKIRT;][]{shvartzvald17a}\footnote{\url{https://exoplanetarchive.ipac.caltech.edu/docs/UKIRTMission.html}}, and 
the Canada-France-Hawaii Telescope \citep[CFHT;][]{zang18}.  
We simultaneously fit the ground-based data and \emph{K2} data and extract 
the \emph{K2} photometry as part of this process.

For microlensing events, the astrophysical difference flux $\tilde{F}_i$
is the \emph{K2} source flux ($F_{s,K2}$) multiplied by 
the magnification $A_i$ with source contribution at baseline ($F_{s,K2}$) 
subtracted:
\begin{equation} \label{eq:flux}
\tilde{F}_i = (A_i - 1) F_{s,K2} .  
\end{equation}
We used MulensModel package \citep{poleski18a_ascl,poleski19a}
to evaluate magnification curves $A_i$.  
Equation~\ref{eq:flux} lacks the baseline flux, $F_{s,K2}+F_{b,K2}$, 
where $F_{b,K2}$ is the blending flux in the \emph{K2} band. 
We have performed a few verification fits with a baseline flux added as a parameter. 
As expected, these fits resulted in zero baseline flux.
This is because the training pixels contain the total flux 
of numerous constant stars, and the additive constant in a model 
(like the baseline flux) is absorbed during the decorrelation process.  
Thus, the MCPM ignores the baseline flux.  
In order to apply the method to other types of variable sources, the definition of $\tilde{F}_i$ (Equation~\ref{eq:flux}) must be modified.  It should be noted that 
we do not need to assume $\tilde{F}_i$ for every epoch; we can limit training 
to a subsample of epochs, train the model, and then extract 
photometry for all epochs.  
This approach can be used to search for short-lasting 
microlensing events or 
planetary anomalies in microlensing light curves. 

\begin{figure}[!h]
\centering
\includegraphics[width=\hsize]{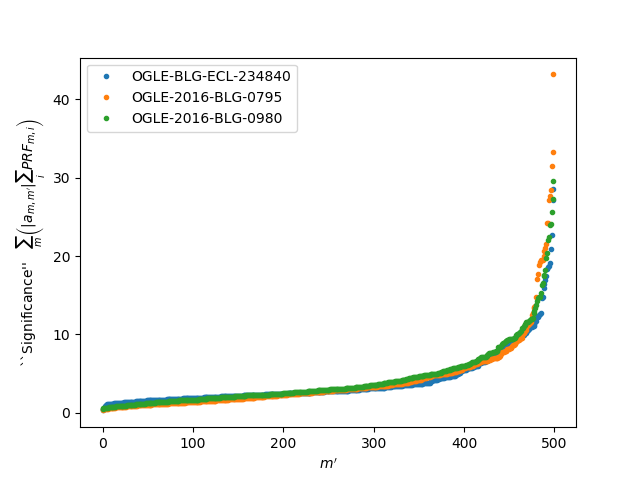}
\caption{
Significance of each individual training pixel to detrending model. 
Higher values of $\sum_{m}\left(\left|a_{m,m'}\right|\sum_i PRF_{m,i}\right)$
correspond to higher significance of a given pixel in constraining the detrending model.
The three colors refer to three targets analyzed in Section~\ref{sec:ex}.  
In each case, about half of the significance is contained in the top $100$ pixels,  
and we use only these pixels in the further analysis.  
\label{fig:aPRF}}
\end{figure}

We first ran the MCPM with $M'=500$ pixels used for training, and 
combined it with $M=4$ pixels used for extracting photometry 
results in as many as $2000$ coefficients to be fit. 
Most of these coefficients are close to zero and do not contribute 
significantly to the trend removal model, see Figure~\ref{fig:aPRF}. 
To reduce the number of model parameters,
we selected $M''=100$ training pixels for which 
$\sum_{m}\left(\left|a_{m,m'}\right|\sum_i PRF_{m,i}\right)$ 
is the largest. Here we used the absolute value of the coefficients 
because both positive and negative values give important information.
Then, we again ran the fitting process using these 
100 pixels for training. 

In our approach, we analyzed \emph{K2} photometry from each subcampaign 
separately, that is, the $a_{m,m'}$ coefficients are different 
for each subcampaign even though the parameters that define 
the astrophysical model are the same.  
For a few events, there are both \emph{K2}C9 and \emph{K2}C11 
data, and the $a_{m,m'}$ coefficients are different in every subcampaign.  

We note that the CPM software by \citet{wang16} 
contained the possibility to include an astrophysical 
model, but it was treated similarly to the signal observed in other pixels, 
meaning, it was multiplied by a coefficient, 
which in turn was subject to regularization.  
This is in contrast to our approach, wherein the astrophysical model 
is subtracted from the target pixel signal before training the model, 
so that the astrophysical model is not subject to regularization.

\subsection{PRF and astrometry}
\label{sec:prf}

One of the key differences between the MCPM and the CPM is the use of the PRF.  
Calculating the fraction of the source flux that falls on a given pixel 
requires a few pieces of information: 
prior knowledge of the source sky coordinates, 
astrometric grid transformation for every epoch, 
the PRF function, and 
an algorithm to interpolate the PRF function.  
For the events detected from the ground, the sky coordinates are known.  
The treatment of events not found from the ground is discussed below.  
The astrometric grid transformation translates 
sky coordinates (right ascension and declination) 
to $(x,y)$ positions on the camera plane.  The \emph{K2} bulge field 
is extremely crowded, and it is difficult to find isolated stars, 
which are required to find the grid transformation.  
For finding the grid transformation we used coordinates from 
the \emph{Gaia} DR1 catalog \citep{gaia16b,gaia16a}.  The \emph{Gaia} passband is similar 
to the \emph{Kepler} passband ($K_p$), which allows us to easily select the brightest 
objects without worrying about the highly variable extinction in the field.  
We measured the positions of the brightest stars using PyKE 
software \citep{pyke3a,pyke3b}, though some very bright stars were not fit properly.  
Thus, the results of PyKE fitting 
were further cleaned based on the inspection of 
the centroid time series plots and astrometric scatter. 
We fit second-order 2D polynomials (12 coefficients in total)
to transform the sky coordinates to $(x,y)$ positions.  
We tried third-order polynomials and found that they did not improve 
the accuracy of the grids significantly.  The dispersion of residuals is 
in the range $0.04-0.11~{\rm pix}$ or $0.16-0.44~{\rm arcsec}$, 
which is sufficient for our purposes. 

To estimate the fraction of the source flux that falls on a given pixel, the MCPM also needs 
the PRF function.  The MCPM uses the \emph{Kepler} PRF function as measured by 
\citet{bryson10} and interpolates it twice.
First, the MCPM uses barycentric interpolation of the five PRFs for every channel 
to account for spatial changes in the PRF. 
Second, the MCPM uses bivariate spline interpolation 
to find the PRF value for every subpixel position.

\subsection{Initial selection of training pixels} 

A selection of $M'$ pixels is necessary for training the model.  
The MCPM selects pixels that are at least 15 pixels away from the target. 
In order to minimize the impact of possible saturated pixels,  
the MCPM removes the pixels that are on the same or neighboring rows as the target, totaling three rows. 
It also removes the pixels that are on the same or neighboring columns as the target.
To further remove the possibility of the overexposed pixels lowering 
the signal, the MCPM excludes the pixels for which the median signal 
(calculated over the whole subcampaign) 
is above $10^5~\mathrm{e^-}\,s^{-1}$. 
We note that some of the training pixels may lie very close to the intrinsically 
varying sources, thus decreasing the power of the model.  
Most importantly, Mira-type variables are bright, have large amplitudes, 
and number nearly 600 inside the superstamp 
\citep{soszynski13b}\footnote{See 
\url{https://www.asc.ohio-state.edu/poleski.1/K2C9_var_stars/} for a list of 
more than $60,000$ variable stars inside the superstamp that was compiled from 
the literature.}.   
At this juncture, we did not remove pixels affected by variable stars 
from the training set.

\subsection{Limitations} 

The MCPM requires prior knowledge of the astrophysical model.  
We do not need to know the exact model, but the prior parameter space 
model must include a model that adequately describes the \emph{K2} data.  
For microlensing events, 
it is possible that the source passed close to a component of the lens system 
as seen by \emph{K2}, but the trajectory seen from the ground did 
not pass this component closely \citep{gould13b,poleski16b,wang18}.  
Identifying such events may be problematic in photometric methods such as ours, which depend on an assumed astrophysical model. 

The MCPM can be run only if we know (or assume) the celestial coordinates 
of the target.  For events not found in the ground-based data, 
we do not know the coordinates, and searching multidimensional parameter 
space ($t_0$, $u_0$, $t_{\rm E}$, $F_{s,K2}$, right ascension, and declination) 
may seem like an extremely computing-intensive task.  
However, there are a few ways of simplifying the calculations.  
First, we may limit the search to short events  
because three independent, high-cadence, ground-based surveys 
\citep[OGLE, KMTNet, and MOA or Microlensing Observations in Astrophysics;][]{bond01}
already searched their \emph{K2}C9 superstamp data and all long events should have been found. 
Second, for the short events and assumed right ascension and declination 
we may exclude a few-day-long part of 
the light curve from training, extract the signal for the whole light curve,
and then check whether the microlensing signal is present in the part excluded 
from training.  To check for the microlensing signal, we only needed to fit 
four parameters: $t_0$, $u_0$, $t_{\rm E}$, and $F_{s,K2}$, which is a simple task. 
It should be noted that for event detection a very coarse 
grid in $u_0$ is enough, for example, \citet{kim18a} used only 0 and 1.  The separation of the event finding process into two independent tasks makes the effort 
more efficient computationally.  
We note that if the part excluded from training is at the beginning or end of 
the campaign, then extracted signals may not be reliable due to changes in 
the spacecraft drift pattern.

\subsection{Fitting process}

For fitting the microlensing model, we used 
the affine-invariant ensemble sampler for Markov chain Monte Carlo 
(called EMCEE) by \citet{foremanmackey13}. 
We defined likelihood:
\begin{equation}
\ln\mathcal{L} = \ln\mathcal{L}_0 - \frac{\chi^2}{2},
\end{equation}
where $\mathcal{L}_0$ is a constant.  The $\chi^2$ for given model is :
\begin{equation}
\chi^2=\sum_i{\left(\frac{F_i-\tilde{F}_i}{\sigma_i}\right)^2} + \sum_j{\sum_k{\left(\frac{F_{j,k}-\left(A_{j,k}F_{s,j}+F_{b,j}\right)}{\sigma_{j,k}}\right)^2}},
\end{equation}
where the first term is the \emph{K2} contribution and 
the second term is the contribution of the ground-based datasets.  
The index $j$ indicates ground-based datasets and $k$ indicates the epochs for 
a given dataset.  The source flux $F_{s,j}$ and blending flux $F_{b,j}$ were 
found via a least-squares fit for a given model and a given dataset.  The microlensing 
model gives $A_{j,k}$ -- a prediction of magnification for a given epoch. 
For each flux measurement $F_{j,k}$ the corresponding uncertainty 
is $\sigma_{j,k}$.  

First, we ran the fitting with $M'=500$ training pixels.  
These pixels contain both pixels in which instrumental trends are
correlated or anticorrelated with trends observed in the target pixels as well as
pixels in which instrumental trends are not correlated with target pixels.
We used the first run to select training pixels that carry useful information,
meaning those that are correlated or anticorrelated with trends observed in the target pixels.
The acceptance rate in this run steadily decreases 
because the chain gets stuck in very narrow minima of $\chi^2$ produced 
by the large number of poorly constrained nuisance parameters. 
Even a small change of model parameters (as compared to parameter uncertainties) results in 
a significant change in $\chi^2$.
The sampler is run typically for $500$ steps because after 
that point, the acceptance rate is very low.  To fully explore the parameter
space, we ran many parallel walkers.
The large number of walkers does not impact the acceptance rate.
For every model, we stored
all $a_{m,m'}$ nuisance parameters.  We calculated the mean of the posterior
distribution of $a_{m,m'}$  using all samples after the first $100$ for each walker.
We note that the acceptance rate problem is worse when the regularization is not strong enough, as in
for small values of regularization constant normalized by the number of training pixels
($\lambda'\equiv\lambda/M'$). 

In the second run, we limited the number of training pixels to $M''=100$ as described above. 
The sampler was run for thousands of steps in order to achieve stable posteriors. 
If a chain with a reasonable acceptance rate was produced, then we reported the results 
from that run.  
We verified that the best-fitting models from
the first run were close to the best-fitting models from the second run.  
If the acceptance rate in the second run was also very small, 
then the sampler was run for the third time. This time, all the $a_{m,m''}$
coefficients were fixed at the mean values as found in the first 500 steps of the second run. 
The third run always produced a reasonable acceptance rate.
We validate the uncertainties resulting from the third run in Section~\ref{sec:795}.

We multiplied the \emph{K2} flux uncertainties by a constant factor that brings 
$\chi^2/\rm{d.o.f.}$ for \emph{K2} close to unity.  
For the determination of the ${\rm d.o.f.}$ we took into account all the $a_{m,m''}$ 
coefficients, of which there are typically $400$. Hence, the $\chi^2$ for 
\emph{K2} data should be $\approx750$ and $\approx1500$ for C9a and C9b, 
respectively. For the examples presented in Section~\ref{sec:ex}, 
the \emph{K2} flux uncertainties were multiplied by $10.0$, $4.48$, and $4.36$, 
respectively, relative to the $f_{m,i}$ uncertainties as reported 
by the \emph{K2} pipeline, which include all expected sources of noise.  

The nuisance parameters $a_{m,m'}$ may be significantly affected by just 
a few epochs with exceptionally large residuals.  The typical scatter of 
the MCPM photometry is in the order of 20 flux units  
($\mathrm{e^-}\,s^{-1}$; the zero-point of photometry is $25~\rm{mag}$). 
Our experience shows that in some cases 
removing epochs with absolute residuals $> 300$ (of which there are typically few) 
improves the model substantially.

The microlensing model fitting is subject to discrete degeneracies and, 
in particular, the satellite parallax measurements are affected by 
the four-fold degeneracy \citep{refsdal66,gould94b}.  In some events this degeneracy 
can be partially or fully broken.  On the other hand, the binary lens events can be affected 
by additional degeneracies \citep{dominik99,skowron11,choi12b}.  
In our approach, each degenerate solution gives slightly different input to 
the minimization process defined by Equations~\ref{eq:main} and~\ref{eq:l2}.   
Hence, the $a_{m,m'}$ coefficients differ and the resulting light curve is different.  
In practice, different degenerate models of 
a given event produce very similar light curves. 

\section{Examples}
\label{sec:ex}

We applied our method to an eclipsing binary and two microlensing events, 
as discussed below. We used photometric data from ground-based 
microlensing surveys. For the OGLE survey data \citep{udalski15b}, the errorbars were rescaled 
following \citet{skowron16a}. For other datasets the errorbars were multiplied 
by a constant that was chosen so that for an initial model fit, a given 
dataset gives $\chi^2/{\rm dof} = 1$. Outlying points were removed from 
the ground-based data. 
The MCPM parameters used in the examples were: 
$M = 4$, $M'=500$, $M''=100$, and $\lambda'=6000$.
We estimated the crowding in the vicinity of targets by calculating the number of 
stars $I<19~{\rm mag}$ within two K2 pixel radii and found: 40, 28, 38, and 39, 
respectively for objects discussed in the following subsections.  Three of these 
numbers are higher than the median for superstamp events of 29. 
The scatter of the \emph{K2} photometry extracted using the MCPM is:
$17$, $18$, and $9.3~{\rm \mathrm{e^-}\,s^{-1}}$ for objects presented in Sections 
\ref{sec:234}, \ref{sec:795}, and \ref{sec:980}, respectively.  
For the two latter objects the \citet{zhu17a} method gives scatter of 
$110$ \citep{zhu17a} and $130~{\rm \mathrm{e^-}\,s^{-1}}$ \citep{zang18}, 
respectively.  Thus, our method has photometric scatter that is smaller by 
a factor of $6.1$ and $14$, respectively.

\subsection{Eclipsing binary OGLE-BLG-ECL-234840}
\label{sec:234}

Eclipses of long-period eclipsing binaries show light curves that are 
similar to inverted microlensing events.  
Unlike microlensing events, however, the eclipses will appear 
the same to an observer in space as they do from 
Earth\footnote{The parallax effect changes the epoch of eclipse, but this effect is negligible \citep{scharf07}.}. 
Hence, we tested our method on the bright, long-period
eclipsing binary OGLE-BLG-ECL-234840 \citep{soszynski16b}.  
The maximum light brightness is $I=13.753~{\rm mag}$ and $V=16.428~{\rm mag}$. 
The orbital period is $369.2~{\rm d}$, and the long-term OGLE light curve 
predicts a primary eclipse at ${\rm HJD} = 2457519.862$, that is, during \emph{K2}C9a. 
We fit Chebyshev polynomial models to phased OGLE $I$- and $V$-band light curves and obtained 
eclipse depths of $\Delta I = 0.398~{\rm mag}$ and $\Delta V = 0.504~{\rm mag}$. 
The $I$- and $V$-band model light curves were transformed to $K_p$-band using the relations presented by 
\citet{zhu17a} and interpolated to the extinction parameters for this line of sight: 
$A_I = 1.42~{\rm mag}$ and $R_I = 1.22~{\rm mag}$ \citep{nataf13b}.
We could not use multiband photometry and relations from CFHT \citep{zang18}  
because the target star falls in the gap between CFHT camera CCD chips. 
The \citet{zhu17a} relations were derived for a single star and are quadratic 
functions of $(V-I)$ color, whereas in eclipsing binaries 
we observe two stars with different intrinsic colors, 
which may cause low-level inaccuracies in the predicted model. 
The resulting $K_p$ light curve has maximum light at $15.426~{\rm mag}$ 
and amplitude of $0.421~{\rm mag}$ 
(in flux space $6752$ and $2255~\mathrm{e^-}\,s^{-1}$, respectively).  We transformed the model curve to 
flux space and normalized it so that maximum light and the faintest eclipse part 
correspond to one and zero, respectively. This model curve is denoted as $f(t)$.  
We calculated $f(t)$ using phase-folded data, hence $f(0) = 0$. 

\begin{figure}
\centering
\includegraphics[width=\hsize]{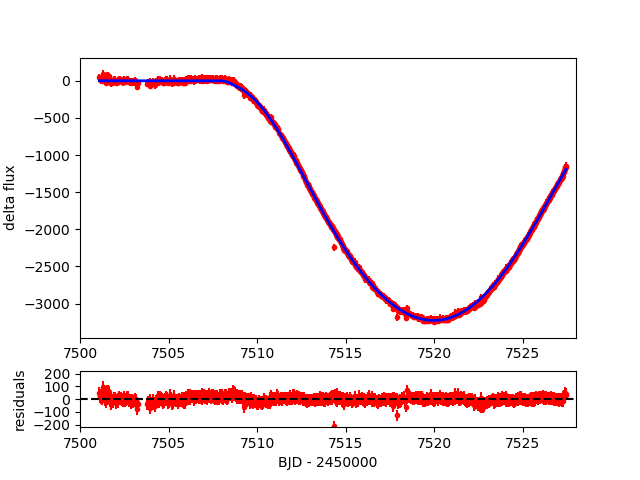}
\caption{\emph{K2} light curve { (represented by red points)} of eclipsing binary OGLE-BLG-ECL-234840 derived using MCPM 
with model light curve (indicated by a blue line) predicted using ground-based data and a three-parameter fit (Equation~\ref{eq:eb}).   
The $y$-axis units are \emph{K2} flux units 
where the photometric zero point of the magnitude scale corresponds to $K_p = 25~{\rm mag}$.  
The lower panel presents residuals of the fit.
\label{fig:eb1}}
\end{figure}

\begin{figure}
\centering
\includegraphics[width=\hsize]{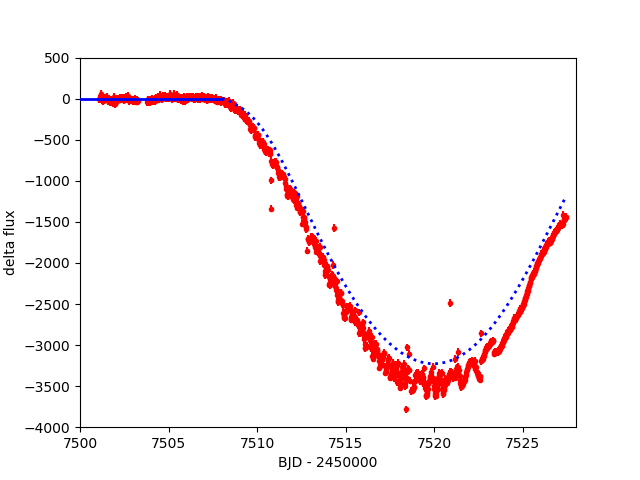}
\caption{\emph{K2} light curve of OGLE-BLG-ECL-234840 derived using MCPM.  
The model was trained using a flat part of the light curve only (indicated by a blue solid line; ${\rm BJD}<2457508$) 
and the full light curve was extracted.
The blue dotted line reproduces the model from Figure~\ref{fig:eb1} for comparison.  
\label{fig:eb2}}
\end{figure}

After preparing the normalized eclipse light curve in $K_p$, 
we applied the MCPM to \emph{K2}C9a data with a flux model defined as
\begin{equation} \label{eq:eb}
 \tilde{F}_i = D f\left(\alpha(t_i-t_0)\right),
\end{equation}
where $D$ is the eclipse depth in flux units, 
$\alpha$ is the eclipse duration stretching factor, and 
$t_0$ is the epoch of the eclipse. 
When combined with the $f(t)$ model from above, the only information that is fixed 
in this approach is the eclipse shape, and the free parameters enable us to 
test the model fitting process. 
We applied the MCPM to \emph{K2} data only and obtained:
$t_0 = 2457519.836\pm0.048$, 
$\alpha = 0.9943\pm0.0064$, 
$D = 3251\pm27$, and 
$\chi^2/{\rm dof} = 739.0/743$.  
The extracted photometry and model light curve are presented in Figure~\ref{fig:eb1}.
The eclipse epoch and the stretching factor are consistent with 
the OGLE predictions within $1\sigma$. 
The measured depth would be consistent with the model light curve if the maximum 
light were $K_p = 15.03~{\rm mag}$, which is $0.40~{\rm mag}$ brighter than 
the prediction.  To verify this discrepancy, we extracted the \emph{K2} light curve 
with training limited to maximum light 
(i.e., ${\rm BJD}<2457508$, $283$ epochs) 
and assumed zero flux during this time.  In this way, we extracted \emph{K2} photometry that is 
independent from our model light curve. We present the resulting data in Figure~\ref{fig:eb2}.  
The eclipse depth differs from the model fitting result, but the difference 
is not large enough to account for the $0.40~{\rm mag}$ discrepancy found above. 
The general shape of the light curve is similar to Figure~\ref{fig:eb1},  
while the scatter of the data is larger.

\subsection{Microlensing event OGLE-2016-BLG-0795}
\label{sec:795}

\begin{figure}
\centering
\includegraphics[width=\hsize]{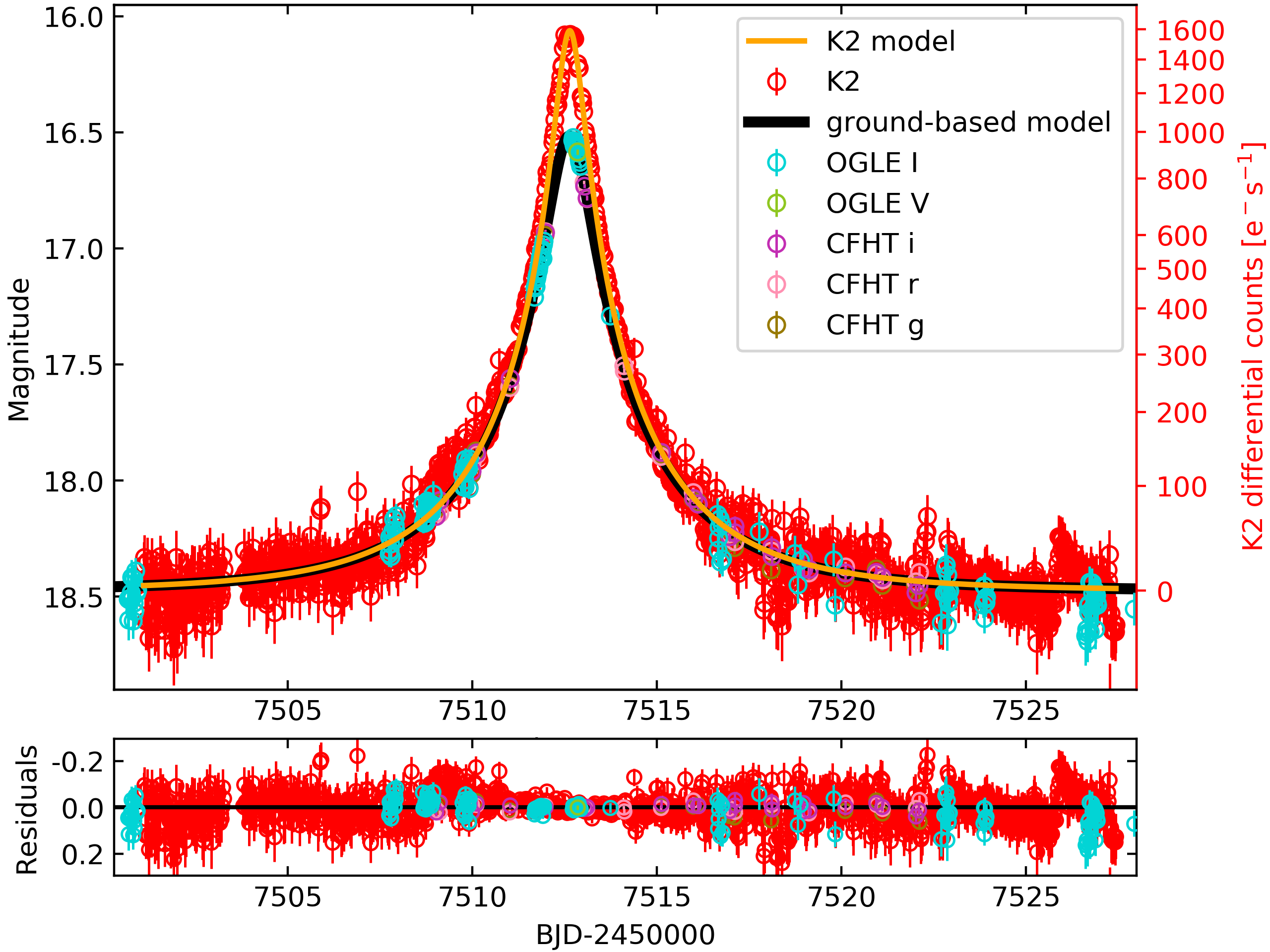}
\caption{MCPM light curve of OGLE-2016-BLG-0795.  
The \emph{K2} light curve (red) has amplitude of 1600 counts, see bottom part
of Figure~\ref{fig:tpf} for raw data.  The right-hand side of the Y axis 
is nonlinear and shows selected \emph{K2} flux values.
Compare to Figure~9 in \citet{zang18}.  
\label{fig:795}}
\end{figure}

We tested our method on the short ($t_{\rm E}=4.5~{\rm d}$) event OGLE-2016-BLG-0795 
that was previously analyzed by \citet{zang18}.  We present the results 
of model fitting in Figure~\ref{fig:795} and Table~\ref{tab:795}.

\begin{sidewaystable*}
\caption{OGLE-2016-BLG-0795 models}
\label{tab:795}
\centering
\begin{tabular}{l r r r r}
\hline\hline
parameter & $(+,+)$\tablefootmark{a} & $(-,-)$ & $(+,-)$ & $(-,+)$ \\ 
\hline                        
$t_0$ & $7512.6276 \pm 0.0037$ & $7512.6276 \pm 0.0036$ & $7512.6269 \pm 0.0035$ & $7512.6283 \pm 0.0037$ \\
$u_0$ & $0.1278 \pm 0.0018$ & $-0.1284 \pm 0.0018$ & $0.1268 \pm 0.0029$ & $-0.1299 \pm 0.0017$ \\
$t_{\rm E}~{\rm [d]}$ & $4.467 \pm 0.022$ & $4.451 \pm 0.019$ & $4.491 \pm 0.067$ & $4.417 \pm 0.021$ \\
$\pi_{{\rm E}, N}$ & $-0.1550 \pm 0.0047$ & $0.1542 \pm 0.0057$ & $-0.746 \pm 0.015$ & $0.7438 \pm 0.0082$ \\
$\pi_{{\rm E}, E}$ & $0.0265 \pm 0.0042$ & $-0.0424 \pm 0.0027$ & $0.1403 \pm 0.0052$ & $-0.1890 \pm 0.0030$ \\
$I_{s}~{\rm [mag]}$ & $19.098 \pm 0.013$ & $19.093 \pm 0.014$ & $19.108 \pm 0.026$ & $19.080 \pm 0.013$ \\
$V_{s}~{\rm [mag]}$ & $20.210 \pm 0.014$ & $20.204 \pm 0.014$ & $20.219 \pm 0.027$ & $20.191 \pm 0.013$ \\
$g_{{\rm PS1}, s}~{\rm [mag]}$ & $20.958 \pm 0.013$ & $20.953 \pm 0.012$ & $20.968 \pm 0.027$ & $20.939 \pm 0.012$ \\
$r_{{\rm PS1}, s}~{\rm [mag]}$ & $20.012 \pm 0.011$ & $20.007 \pm 0.011$ & $20.020 \pm 0.026$ & $19.994 \pm 0.011$ \\
$i_{{\rm PS1}, s}~{\rm [mag]}$ & $19.466 \pm 0.012$ & $19.461 \pm 0.011$ & $19.475 \pm 0.026$ & $19.448 \pm 0.011$ \\
$K_{p,s}$ & $19.589 \pm 0.012$ & $19.591 \pm 0.011$ & $19.583 \pm 0.019$ & $19.594 \pm 0.012$ \\
\hline
$\chi^2/{\rm dof}$ & $2102.64/2016$ & $2100.91/2016$ & $2106.48/2016$ & $2103.56/2016$ \\
\end{tabular}
\tablefoot{
Baseline brightness is $18.771~{\rm mag}$ in $I$ band
and $19.822~{\rm mag}$ in $V$ band \citep{szymanski11}.  
In each case \emph{K2} peak time and impact parameters are $t_{0,K2}=7512.642$ and 
$u_{0,K2}=\pm0.084$.  Corresponding parameters for \citet{zang18} models are:
$(7512.724, 0.126)$, 
$(7512.724, -0.128)$, 
$(7512.724, -0.132)$, and 
$(7512.745, 0.132)$ in the order as in the table above. \\ 
\tablefoottext{a}{The two signs indicate $u_0$ signs as seen for Earth and \emph{K2}, respectively.  
See also Figure~2 of \citet{gould94c}.}
}
\end{sidewaystable*}

For all plots of microlensing event light curves we scaled all the data to 
a common photometric system so that data from different telescopes and in 
different passbands can be compared to just one model curve 
(or two if satellite data are used). The standard method is to first translate 
the measured flux $F$ to the observed magnification space: $A = (F-F_b)/F_s$, 
where $F_b$ and $F_s$ are the blending and source fluxes for a given 
photometric system. Second, the observed magnification is 
translated to the photometric system of a reference dataset 
(OGLE $I$-band in our case): 
$F_{\rm ref} = A F_{s, {\rm ref}} + F_{b, {\rm ref}}$, 
where $F_{b, {\rm ref}}$ and $F_{s, {\rm ref}}$ are the blending and source 
fluxes for the reference dataset.  
All source and  blending fluxes are found via linear regression.  

Our \emph{K2} photometry differs from that of \citet{zang18}, 
which was extracted using the \citet{zhu17a} approach.  
Our four models have peak \emph{K2} magnification ($A_0$) in the range $11.9-12.0$, 
or peak $(A_0-1)F_{s,K2}$ of $1600$. 
The four \citet{zang18} models have peak magnification in the range $7.6-8.0$ 
and corresponding $(A_0-1)F_{s,K2}$ from $550-590$.  
This is a significant difference, and we try to verify which model is correct 
by running the MCPM on the \citet{zang18} models and 
with the MCPM trained on the nearly flat part of the light curve, 
meaning the union of ${\rm BJD} < 2457510$ and ${\rm BJD} > 2457515$.
All four \citet{zang18} models result in peak fluxes of $1550$, 
which is very close to the results from the MCPM fit on the whole light curve.
We conclude that in our framework of decorrelating \emph{K2} signals against 
signals in other pixels, the \citet{zang18} model is inconsistent with 
the photometry extracted using the MCPM.  
We can imagine that the models that were fit using the two methods differ 
because the MCPM is overfitting.  To test if this is the case, we ran the fit 
with a smaller number of training pixels. We found that with $M'=6$ 
(i.e., there are just 24 nuisance parameters)
the microlensing signal is still extracted and consistent with 
Table~\ref{tab:795}, though with larger uncertainties.  
We obtained consistent results even with $M'=5$ and a single additional change 
($\lambda' = 500$).  To sum up, our tests do not show signs of overfitting.

The results presented in Table~\ref{tab:795} were obtained 
from the second run of the sampler 
with $M''=100$ and $a_{m,m''}$ as free parameters. We then ran 
the sampler a third time (i.e., $a_{m,m''}$ are fixed) 
for each of the models and compared the resulting 
posterior statistics. The uncertainties are on average $1.6$ times larger 
when compared to the second run.  The largest ratios are for 
$t_E$ and are up to $3.5$ times higher.  
The mean values from both runs are consistent when compared to the uncertainties 
from the third run.
We conclude that the third run returns posteriors that are consistent with 
the second run, though with larger uncertainties.

We also performed additional fits to check how the \emph{K2} source flux 
constraints affect the MCPM fitting results.  We use the predictions 
made by \citet{zhu17a} where $(K_p-I)$ was parameterized as a function 
of $(V-I)$ and the extinction parameters at a given sight-line. 
We used $A_I = 1.04~{\rm mag}$ and $E(V-I) = 0.88~{\rm mag}$ \citep{nataf13b}. 
The $(V-I)$ color was estimated using OGLE data. 
When the \citet{zhu17a} calibration is applied to the results presented in 
Table~\ref{tab:795}, the predicted $(K_{p,s}-I_s)$ color is larger by 
$0.366~{\rm mag}$ than the MCPM fitting result. 
In our fitting routine, we added the $\chi^2$ penalty: 
$\left((K_{p,s}-\tilde{K}_{p,s})/0.02~{\rm mag}\right)^2$, 
where $\tilde{K}_{p,s}$ is the \emph{K2} brightness predicted using 
the \citet{zhu17a} calibration. 
The $\chi^2$ penalty was calculated for every model. 
The resulting fits have parameters that are significantly different from  
those presented in Table~\ref{tab:795} and have $\chi^2$ higher by $130$. 
We conclude that the MCPM method gives results inconsistent with the \citet{zhu17a} calibration. 

The measurement of $\mbox{\boldmath$\pi$}_{\rm E}$ allows an estimation of  the relative 
heliocentric lens-source velocity projected on the observer plane 
($\mbox{\boldmath${\it \tilde{v}_{\rm hel}}$}$):
\begin{equation}
\mbox{\boldmath${\it \tilde{v}_{\rm hel}}$} = \mbox{\boldmath${\it \tilde{v}_{\rm geo}}$} + \mbox{\boldmath${\it v_{\oplus,\perp}}$}
\end{equation}
where
\begin{equation}
\mbox{\boldmath${\it \tilde{v}_{\rm geo}}$} = \frac{\mbox{\boldmath$\pi$}_{\rm E}}{\pi_{\rm E}^2} \frac{\rm AU}{t_{\rm E}},
\end{equation}
and $\mbox{\boldmath${\it v_{\oplus,\perp}}$}$ is the velocity of Earth 
at $t_{0,\oplus}$ projected on the plane of the sky. 
For OGLE-2016-BLG-0795, the preferred solutions are $(+,+)$ and $(-,-)$ 
\citep[by { the} so-called Rich argument, see][]{zang18}.  The projected heliocentric velocities are (N, E):
$(-2428\pm104, 435\pm68)~{\rm km\,s^{-1}}$ for the $(+,+)$ solution and  
$(2347\pm120, -645\pm47)~{\rm km\,s^{-1}}$ for the $(-,-)$ solution.  
The relative lens-source heliocentric proper motion ($\mu_{\rm hel}$) 
is related to $\tilde{v}_{\rm hel}$ via: 
$\tilde{v}_{\rm hel} = {\rm AU}\mu_{\rm hel}\pi_{\rm rel}^{-1},$ 
which can be rewritten as:
\begin{equation} \label{eq:pirel}
\pi_{\rm rel} = {0.01~{\rm mas}}\frac{\mu_{\rm hel}}{5~{\rm mas\,yr^{-1}}}\left(\frac{\tilde{v}_{\rm hel}}{2400~{\rm km\,s^{-1}}}\right)^{-1}.
\end{equation}
This suggests that $\pi_{\rm rel}$ is small, hence, the lens is close to the source. 
The most likely interpretation is that the lens is located in the Galactic bulge.

\subsection{Microlensing event OGLE-2016-BLG-0980}
\label{sec:980}

\begin{figure}
\centering
\includegraphics[width=\hsize]{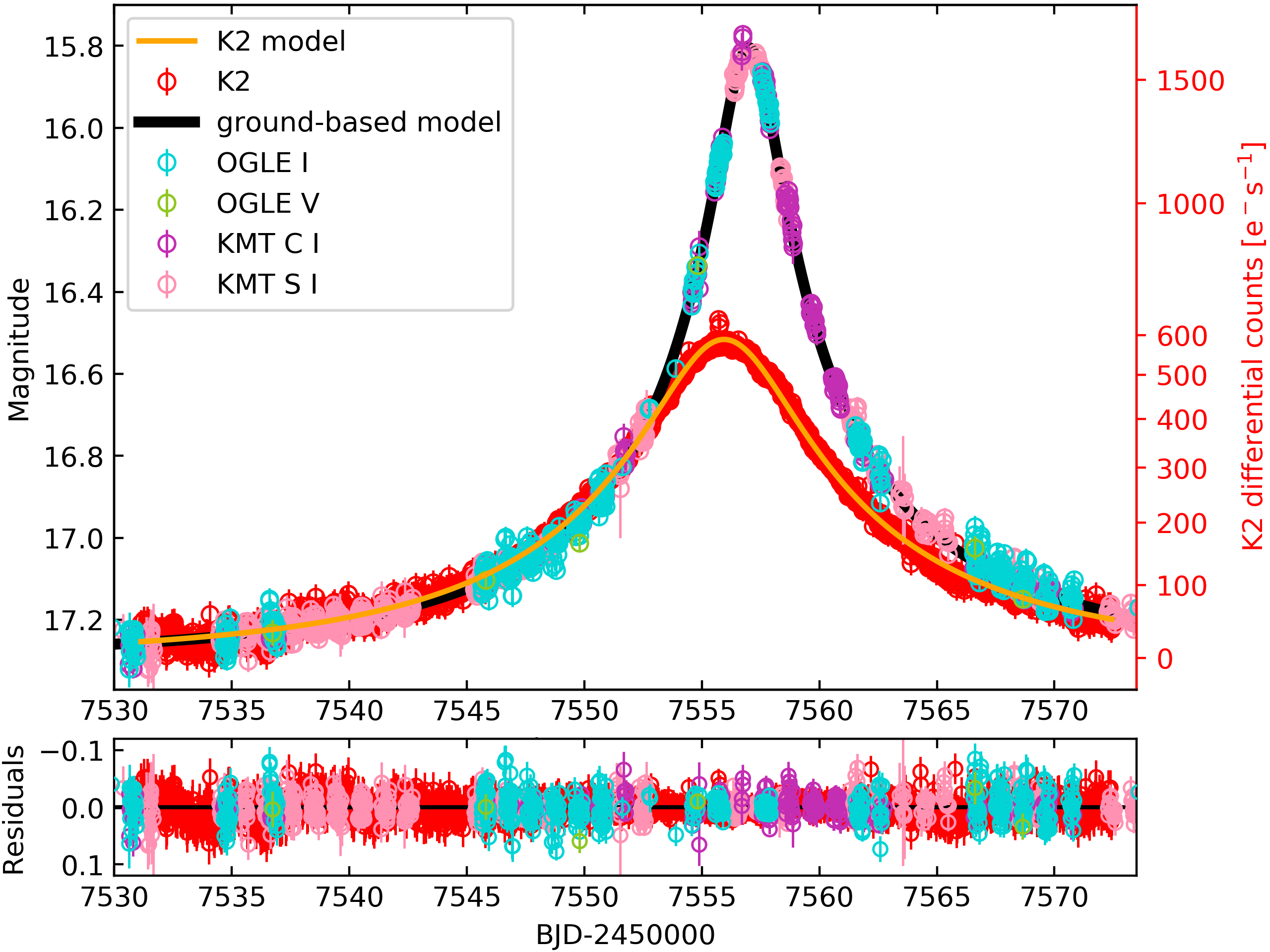}
\caption{MCPM light curve of OGLE-2016-BLG-0980.   
Compare to Figure~4 in \citet{zhu17a}.  
\label{fig:980}}
\end{figure}

OGLE-2016-BLG-0980 was first modeled by \citet{zhu17a}. 
The OGLE $I$-band data show a slight dependence on airmass so we removed this trend from the data.  
We also used KMTNet data \citep{kim16,kim18b} from the Cerro Tollo Inter-American Observatory (Chile; designated C) 
and the South African Astronomical Observatory (South Africa; designated S). 
The KMTNet data from the Siding Spring Observatory (Australia; designated A) are noisy and do not contribute significantly to constraining the model; we therefore did not include them. 
We present the results of model fitting in Figure~\ref{fig:980} and Table~\ref{tab:980}. 
As compared to \citet{zhu17a} results, we note differences in $t_0$, $u_0$, and $t_{\rm E}$ that are caused by 
the fact that we detrended against airmass, whereas \citet{zhu17a} did not.
The parallax results are statistically different, but the differences are small. 
The differences are comparable to the parallax uncertainties measured using 
the annual parallax effect for other events.  
The scatter of the data is significantly smaller in the MCPM reduction 
as compared to \citet{zhu17a}.  

From Table~\ref{tab:980} we see that the $u_0>0$ solution is clearly preferred. 
The projected heliocentric velocity is $(520\pm13, 264.0\pm5.5)~{\rm km\,s^{-1}}$. 
Combining this value with Equation~\ref{eq:pirel}, we see that 
the lens and source parallaxes differ significantly, which 
suggests that the lens is located in the Galactic disk. 

\begin{table}
\caption{OGLE-2016-BLG-0980 models}
\label{tab:980}
\centering
\begin{tabular}{l r r}
\hline\hline
parameter & $u_0>0$ model & $u_0<0$ model \\ 
\hline                        
$t_0$ & $7556.9980 \pm 0.0029$ & $7556.9974 \pm 0.0029$ \\
$u_0$ & $0.06990 \pm 0.00096$ & $-0.06920 \pm 0.00096$ \\
$t_{\rm E}~{\rm [d]}$ & $18.45 \pm 0.18$ & $18.59 \pm 0.19$ \\
$\pi_{{\rm E}, N}$ & $0.1499 \pm 0.0024$ & $-0.1515 \pm 0.0024$ \\
$\pi_{{\rm E}, E}$ & $0.0677 \pm 0.0011$ & $0.0791 \pm 0.0012$ \\
$I_{s}~{\rm [mag]}$ & $18.870 \pm 0.015$ & $18.881 \pm 0.015$ \\
$V_{s}~{\rm [mag]}$ & $20.344 \pm 0.014$ & $20.355 \pm 0.014$ \\
$K_{p,s}~{\rm [mag]}$ & $19.765 \pm 0.018$ & $19.711 \pm 0.018$ \\
\hline
$\chi^2/{\rm dof}$ & $7350.11/6324$ & $7447.17/6324$ \\
\hline                                  
\end{tabular}
\tablefoot{
Baseline brightness is $17.229~{\rm mag}$ in  $I$ band
and $18.551~{\rm mag}$ in $V$ band \citep{szymanski11}. 
The \emph{K2} peak time and impact parameter 
$(t_{0,K2}, u_{0,K2})$ for the above models are 
$(7555.945, 0.176)$ and $(7555.945, -0.181)$, respectively.  
For the \citet{zhu17a} model and without a color constraint, these 
parameters are $(7556.027, 0.142)$. 
}
\end{table}

\subsection{Microlensing event MOA-2016-BLG-290}

\citet{zhu17c} analyzed photometry of the microlensing event MOA-2016-BLG-290 from three 
different locations in the Solar System: Earth, \emph{K2}, and \emph{Spitzer}. 
They fit a parallax point-source point-lens model to the ground-based data and extracted 
the \emph{K2} photometry using the \citet{zhu17a} method.  
The ground-based data were from MOA and OGLE surveys, and for OGLE data, zero blending flux was assumed.  
Four degenerate solutions were 
found and were further verified using twelve epochs of \emph{Spitzer} 
photometry that cover only the falling part of the light curve.  
\citet{zhu17c} predicted \emph{Spitzer} 
light curves and fit the source and blending fluxes for \emph{Spitzer} data. 
Two of the four solutions have \emph{Spitzer} source fluxes that are consistent 
with the prediction based on a color-color relation derived using nearby stars. 
This allowed \citet{zhu17c} to break the four-fold $\mbox{\boldmath$\pi$}_{\rm E}$ degeneracy.  
One can also consider the source flux consistency as a strong argument showing the reliability 
of the \citet{zhu17a} method.

We tried to reduce \emph{K2} data for MOA-2016-BLG-290, 
but quickly arrived at a problem with fitting the ground-based data alone: 
The best point-source point-lens fit requires significant negative blending flux. 
The negative blending flux can be caused by an incorrect model, or 
can be naturally produced if the event occurs on a ``hole'' in an otherwise 
roughly uniform background of bulge sources \citep{park04}. As an example, 
\citet{yee15a} suggested that the blending flux in $I$-band: 
$F_{b,I} > -0.2$ (where the zero-point corresponds to $18~\rm{mag}$) 
can be explained this way. In other words, the uniform background 
would be composed of $I=19.75~\rm{mag}$ stars. 
In the case of MOA-2016-BLG-290, we fit $I$-band data from OGLE and two 
overlapping fields from both KMT A and KMT S.  The fit results in 
$\chi^2/\rm{dof} = 7569.9/7843$ and the blending flux of $-3.15\pm0.50$ 
This is equivalent to a hole in a uniform background of $16.76~\rm{mag}$ stars,
which is unlikely given the source density and luminosity function of 
stars toward the bulge.  Thus, we conclude that the inferred blending flux is 
significantly more negative than can be explained using the \citet{park04} 
interpretation of a hole in a uniform background of stars.
After adding a prior $F_{b,I,{\rm OGLE}} > -0.2$, we obtained 
$\chi^2/\rm{dof} = 7612.1/7843$ and the blending flux of $-0.158^{+0.068}_{-0.032}$.  
We conclude that the negative blending flux cannot be explained as a hole in a uniform background 
of stars, nor can it be explained by
systematic effects in the photometry because it is present in a joint fit to data 
from three telescopes.  Hence, it is most likely 
caused by some second-order effect, such as: a finite source, a binary source, 
xallarap, or a binary lens.  

After considering the uniform finite-source models \citep{lee09}, we
found $\chi^2/\rm{dof} = 7561.2/7842$, $\rho=0.471\pm0.075$, 
and the blending flux in this case is $-1.42\pm0.48$.  
Thus, a single additional parameter gives an improvement of $\Delta\chi^2 = 8.7$. 
In this case, the baseline object is about $1~{\rm mag}$ brighter than the red clump
and hence would have an angular source size ($\theta_{\star}$) of roughly 
$10~{\rm \mu as}$.  In this finite-source model, 
we can estimate physical properties: $\theta_{\rm E} = \theta_{\star} / \rho \approx 0.022~{\rm mas}$ and 
the relative lens-source geocentric proper motion 
$\mu_{\rm geo} = \theta_{\rm E}/t_{\rm E} \approx 1.3~{\rm mas\,yr^{-1}}$.  
Our preliminary extraction of \emph{K2} photometry suggests 
$t_{0,K2}\approx7553.105$ and $u_{0,K2}\approx u_{0,\oplus}$, that is, 
$\pi_{\rm E} \approx 0.17$.  If these values were correct, then we would infer a lens mass of 
$M = 16\left(\frac{\pi_{\rm E}}{0.17}\right)^{-1}~{\rm M_{\rm Jup}}$.  
A priori probability of such a small $\theta_{\rm E}$ value seems to be small. 
We also fit the binary-source model with the $F_{b,I}$ prior, 
resulting in $\chi^2/\rm{dof} = 7564.3/7840$ or $\Delta\chi^2=47.8$ for 
three additional parameters.  
A binary companion to such a bright source also seems unlikely. 

The significantly negative blending flux suggests that the correct model for 
MOA-2016-BLG-290 has not yet been found.  In principle, the agreement between 
the predicted and fit \emph{Spitzer} source fluxes for 
the small-parallax solutions in \citet{zhu17a} 
would argue that the their solutions were correct, however this agreement 
could be 
coincidental.  We conclude that a more in-depth analysis of this event is needed, 
which is beyond the scope of this paper.

\section{Flux calibration}
\label{sec:flux}

In the previous section, we saw that there are significant differences 
between the flux we measured and the $K_p$ magnitudes predicted using 
the \citet{zhu17a} and \citet{zang18} relations.  The calibration of fluxes 
is important for microlensing events because microlensing models predict 
magnification, which is not a directly measured quantity.  We can measure 
the magnified flux and then estimate the magnification when the source flux is 
constrained. However, the precise knowledge of $K_p$ magnitudes is not needed for 
many other applications of \emph{Kepler} and \emph{K2} data.  
Most importantly, in the case of planetary transits, 
the fundamental quantity is the relative transit depth.

The best approach to constrain the source flux is to directly measure fluxes 
in the \emph{K2} data for a number of isolated stars and then check how these fluxes 
relate to predictions. The number of truly isolated \emph{K2}C9 superstamp 
stars is very small, because the superstamp was selected to cover high-stellar 
density regions. The most isolated superstamp stars are in the regions of
particularly high extinction and these stars are in front of most of 
the dust.  Hence, these objects 
are relatively nearby main-sequence stars and their spectral energy 
distributions are different from the microlensing sources.  Thus, we searched 
for isolated red giants and supergiants, which are then less isolated than 
the nearby main-sequence objects.  We searched the OGLE-IV photometric catalog 
calibrated to the standard system because it covers the superstamp fully
(except small gaps between CCD chips) and 
the OGLE-IV data are available for almost all \emph{K2}C9 microlensing events.  
To select isolated stars for our test, we applied a few constraints: 
1) the star is not overexposed in the OGLE $I$-band data, meaning, $I>12~{\rm mag}$; 
2) the other stars contribute relatively little flux in the OGLE $I$-band data; 
3) there are no overexposed pixels in the \emph{K2}C9 data; 
4) the position of the star on the extinction-corrected \citep{nataf13b} 
color-magnitude diagram is consistent with red giants or supergiants; and 
5) the star is relatively constant ($\sigma_I < 0.025~{\rm mag}$ in the OGLE data).

In order to estimate \emph{K2} brightness, we first estimated background 
flux for the given target and subcampaign.  We selected a square of $5\times5$ pixels centered 
on the target and sorted these pixels according to the PRF contributions from 
the target.  We ignored the top ten pixels and for each remaining pixel we 
took the median flux measured over a subcampaign and then took the median 
over the pixels, which results in the background estimate for a given subcampaign ($b$).  
To estimate the target flux, we assumed the background is properly measured 
which leads to the following equation for the flux estimate ($F'_i$):
\begin{equation}
F'_i = \frac{\sum_l PRF_{l,i}\left(f_{l,i}-b\right)}{\sum_l PRF_{l,i}^2}.
\end{equation}
The index $l$ indicates five pixels with the highest PRF contribution 
from the given target.  The resulting photometry shows trends with 
spacecraft pointing and changes between the two subcampaigns.  
We report median fluxes and their uncertainties in Table~\ref{tab:flux}.  
Table~\ref{tab:flux} also gives $K_p$ brightness that was estimated using 
the \citet{zhu17a} calibration and OGLE data.  The last column of the table gives the
difference between predicted and measured values.  These differences are 
much larger than the uncertainties of measured flux.

There is a significant scatter of differences between measured and 
predicted $K_p$ magnitudes, however, all differences have the same sign.  
The mean difference is $\langle\Delta K_p\rangle = 0.47\pm0.13~{\rm mag}$ and 
the measured value is brighter than the prediction.  
This difference is significant and is most probably caused by 
a combination of two effects:
flat-field errors (which are largest at the edges of camera field of view) 
and the inaccurately measured zero-point of the $K_p$ band. 

In Section~\ref{sec:ex} we have discussed the MCPM photometry for three targets. 
In each case, the source was found to be brighter than the prediction 
 based on the \citet{zhu17a} calibration of: $0.40~{\rm mag}$, 
$0.37~{\rm mag}$, and $0.19~{\rm mag}$, respectively. 
The first two of these values are close to $\langle\Delta K_p\rangle,$ 
and the third one is $2.2\sigma$ away. We conclude that the MCPM fitting results 
are consistent with the \citet{zhu17a} calibration after correcting for 
$\langle\Delta K_p\rangle$.

\begin{sidewaystable*}
\caption{Photometry of isolated stars}
\label{tab:flux}
\centering
\begin{tabular}{r r r r r r r r r}
\hline\hline
No. & RA (J2000) & Dec (J2000) & \emph{K2}C9 & $I$ & $(V-I)$ & $K^{\rm pred}_p$ & $K^{\rm measure}_p$ & $\Delta K_p$ \\
 & & & channel & ${\rm [mag]}$ & ${\rm [mag]}$ & ${\rm [mag]}$ & ${\rm [mag]}$ & ${\rm [mag]}$ \\
\hline
1 & 18 06 07.66 & $-$27 11 50.68 & 49 & 12.203 & 1.818 & 13.482 & $13.011^{+0.028}_{-0.045}$ & 0.471 \\
2 & 18 04 55.98 & $-$27 06 46.37 & 49 & 12.396 & 2.594 & 14.046 & $13.459^{+0.041}_{-0.047}$ & 0.587 \\
3 & 18 03 23.63 & $-$26 57 12.35 & 49 & 12.573 & 3.460 & 14.499 & $13.959^{+0.042}_{-0.051}$ & 0.540 \\
4 & 17 59 07.58 & $-$28 36 15.16 & 31 & 12.484 & 2.068 & 13.894 & $13.714^{+0.093}_{-0.146}$ & 0.180 \\
5 & 18 05 09.36 & $-$27 15 52.74 & 49 & 12.688 & 1.714 & 13.910 & $13.473^{+0.033}_{-0.053}$ & 0.437 \\
6 & 18 04 51.97 & $-$27 21 14.44 & 49 & 12.546 & 3.309 & 14.481 & $14.042^{+0.084}_{-0.070}$ & 0.439 \\
7 & 18 03 14.89 & $-$27 38 04.09 & 52 & 12.645 & 2.819 & 14.398 & $13.955^{+0.086}_{-0.055}$ & 0.443 \\
8 & 18 04 03.23 & $-$27 50 52.66 & 52 & 12.642 & 3.927 & 14.761 & $14.096^{+0.044}_{-0.036}$ & 0.665 \\
9 & 18 03 58.43 & $-$27 58 18.26 & 52 & 12.536 & 2.349 & 14.083 & $13.579^{+0.046}_{-0.023}$ & 0.504 \\
10 & 18 05 18.04 & $-$27 58 09.62 & 52 & 12.414 & 2.986 & 14.233 & $13.824^{+0.057}_{-0.051}$ & 0.409 \\
\hline
\end{tabular}
\tablefoot{
Mean $I$-band brightness and $(V-I)$ colors come from OGLE survey.  
$K^{\rm pred}_p$ is the brightness predicted using the \citet{zhu17a} calibration, 
and $K^{\rm measure}_p$ is the median measured brightness with 
uncertainties defined by 0.16 and 0.84 percentiles.  The last column gives
$\Delta K_p = K^{\rm pred}_p-K^{\rm measure}_p$.
}
\end{sidewaystable*}

\section{Summary}
\label{sec:end}

We have presented a novel method for extracting photometry 
from highly blended \emph{K2} data.  
The method combines the PRF photometry with a data-driven model 
that removes instrumental effects. 
The removal of instrumental effects depends on model training that can 
be done on the full light curve or on only a part of the data, 
enabling an efficient search for very short 
events and short (e.g., planetary) anomalies. 
Our method of the removal of systematic trends in the photometry 
is designed in a way that preserves the intrinsic 
astrophysical signal.

We found that the MCPM produces photometry that is an order of magnitude more 
precise than the photometry extracted using the \citet{zhu17a} method and that some of 
the results differ.  
Both methods are based on the methods developed previously for the less crowded \emph{K2} campaigns. 
Both methods use an astrophysical model to decorrelate instrumental noise, 
but they decorrelate against different pieces of information. 
An important aspect of the MCPM is the direct use of the PRF, 
which is not employed in the \citet{zhu17a} method.  
In addition to running the MCPM on two microlensing events, we also tested the MCPM on an eclipse 
of a long-period binary.  The inverted shape of this eclipse is comparable to the shape 
of a microlensing event light curve.  The epoch of the eclipse and its length 
were measured to be consistent with the prediction based on the ground-based data.  
We measured the eclipse depth using two approaches within the MCPM. These approaches are roughly consistent 
with each other, but in order to be consistent with the predicted eclipse depth, require the object to be brighter by $0.40~{\rm mag}$ 
than the prediction based on the \citet{zhu17a} calibration. 
We have also extracted photometry for ten bright and isolated stars 
using a method that is similar to PRF fitting.  In all cases, we find 
these stars to be brighter than the \citet{zhu17a} calibration, with 
the mean difference similar to the brightness difference found for 
the eclipsing binary. We conclude that the \citet{zhu17a} calibration requires 
an additive constant.  The \citet{zang18} and the MCPM results for OGLE-2016-BLG-0795 
differ and we ran a number of tests to verify the MCPM results.  These tests 
result in parameters that are very close to our main solution.  We also find 
that \emph{K2} flux calibration is consistent with the correction found on 
isolated stars.  We conclude, that the MCPM results are favored 
over the \citet{zang18} results.

We distribute the MCPM software via:
\begin{center}
\url{https://github.com/CPM-project/MCPM} 
\end{center}
Our astrometric transformations are distributed together with the MCPM code.

\begin{acknowledgements}
We thank David Hogg, Chelsea Huang, Przemek Mr\'oz, Andrew Vanderburg, Dun Wang, 
Weicheng Zang, and Wei Zhu for consultation.  
This work was partly supported by NASA grant NNX17AF72G to R.P. 
OGLE project has received funding from the Polish National Science
Center, grant MAESTRO 2014/14/A/ST9/00121 to A.U.  
The work by C.R. was supported by an appointment to the NASA Postdoctoral Program 
at the Goddard Space Flight Center, administered by USRA through a contract with NASA.  
Work by A.G. was supported by AST-1516842 from the US NSF 
and by JPL grant 1500811.
A.G. received support from the European Research Council under 
the European Union's Seventh Framework Programme (FP 7) 
ERC Grant Agreement n. [321035].
This research has made use of the KMTNet system operated by the Korea
Astronomy and Space Science Institute (KASI) and the data were obtained at
three host sites of CTIO in Chile, SAAO in South Africa, and SSO in
Australia. 
Some of the data presented in this paper were obtained from 
the Mikulski Archive for Space Telescopes (MAST). STScI is operated by 
the Association of Universities for Research in Astronomy, Inc., under NASA contract NAS5-26555.  
This paper includes data collected by the Kepler mission. 
Funding for the Kepler mission is provided by the NASA Science Mission directorate. 

\end{acknowledgements}

\bibliographystyle{aa}
\bibliography{paper}

\end{document}